\documentstyle[11pt,aaspp4]{article}

\received{}
\revised{}
\accepted{}

\begin{document}


\title{Discovery of the Low-Redshift Optical Afterglow of GRB\,011121 and Its
       Progenitor Supernova 2001ke\altaffilmark{1}}

\author{P.~M.~Garnavich\altaffilmark{2}, 
K.~Z.~Stanek\altaffilmark{3},
L.~Wyrzykowski\altaffilmark{4},
L.~Infante\altaffilmark{5},
E.~Bendek\altaffilmark{5}, 
D.~Bersier\altaffilmark{3},
S.~T.~Holland\altaffilmark{2}, 
S.~Jha\altaffilmark{3},
T.~Matheson\altaffilmark{3}, 
R.~P.~Kirshner\altaffilmark{3},
K.~Krisciunas\altaffilmark{6}, 
M.~M.~Phillips\altaffilmark{7},
R.~G.~Carlberg\altaffilmark{8}}

\author{\small\tt e-mail: pgarnavi@miranda.phys.nd.edu,
       kstanek@cfa.harvard.edu, wyrzykow@astrouw.edu.pl,
       linfante@astro.puc.cl, ebendek@astro.puc.cl,
       dbersier@cfa.harvard.edu, sholland@nd.edu,
       saurabh@cfa.harvard.edu, tmatheson@cfa.harvard.edu,
       rkirshner@cfa.harvard.edu, kkrisciunas@noao.edu, mmp@lco.cl,
       carlberg@astro.utoronto.ca}
 
\altaffiltext{1}{\vspace{-0.15cm}Based on data from the OGLE 1.3m and
the Magellan 6.5m Baade telescopes and the Hubble Space Telescope.}

\altaffiltext{2}{\vspace{-0.15cm}Department of Physics, University of Notre Dame,
                 Notre Dame, IN 46556--5670, U.S.A.}

\altaffiltext{3}{\vspace{-0.15cm}Harvard-Smithsonian Center for Astrophysics,
		 60 Garden Street, 
                 Cambridge, MA 02138,
                 U.S.A.} 

\altaffiltext{4}{\vspace{-0.15cm}Warsaw University Observatory, 
                 Al.~Ujazdowskie 4, 
                 00--478 Warszawa, Poland}

\altaffiltext{5}{\vspace{-0.15cm}Pontificia Universidad Cat{\'o}lica de Chile, 
                 Casilla 306, Santiago, 22, Chile}

\altaffiltext{6}{\vspace{-0.15cm}Cerro Tololo Inter-American Observatory,
                 950 N. Cherry Ave., 
                 Tucson, AZ 85719, U.S.A.}

\altaffiltext{7}{\vspace{-0.15cm}Carnegie Institution of Washington,
                 Las Campanas Observatory,
                 Casilla 601,
                 La Serena, Chile}

\altaffiltext{8}{\vspace{-0.15cm}University of Toronto,
                 Dept. of Astronomy,
                 60 Saint George St., 
                 Toronto, ON M5S 3H8 Canada}

\begin{abstract}

\vspace{-0.2cm} 
\baselineskip=13pt

We present the discovery and follow-up observations of the afterglow
of the Gamma-Ray Burst (GRB) 011121 and its associated supernova
SN~2001ke. Images were obtained with the
OGLE 1.3m telescope in $BVRI$ passbands, starting $10.3\;$hours after
the burst.  The temporal analysis of our early data indicates a steep
decay, independent of wavelength with $F_\nu\propto t^{-1.72\pm
0.05}$. There is no evidence for a break in the light curve earlier
than 2.5 days after the burst. The spectral energy distribution
determined from the early broad-band photometry is a power-law with
$F_\nu\propto \nu^{-0.66\pm 0.13}$ after correcting for a large
reddening.  Spectra, obtained with the Magellan 6.5m Baade telescope,
reveal narrow emission lines from the host galaxy which provide a
redshift of $z=0.362\pm 0.001$ to the GRB.  We also present late $R$ and
$J$-band observations of the afterglow $\sim 7-17\;$days after the
burst.  The late-time photometry shows a large deviation from the
initial decline and our data combined with Hubble Space Telescope
photometry provide strong evidence for a supernova peaking about
12 rest-frame days after the GRB. The first spectrum ever obtained
of a GRB supernova at cosmological distance revealed a blue continuum.
SN~2001ke was more blue near maximum than SN~1998bw and faded more
quickly which demonstrates that a range of properties
are possible in supernovae which generate GRBs. The blue color is
consistent with a supernova interacting with circumstellar gas
and this progenitor wind is also evident in the optical afterglow. 
This is the best evidence to date that classical, long gamma-ray
bursts are generated by core-collapse supernovae.

\end{abstract}

\vspace{-0.4cm}


\section{INTRODUCTION}

The origin of gamma-ray bursts (GRB) has been a mystery since their
discovery in the 1960's.  It has only been since the BeppoSAX
satellite (Boella et al.~1997) began providing rapid, accurate
localization of several bursts per year has it been possible to study
these events and their afterglows in detail. Optical observations of
afterglows (e.g. GRB 970228: Groot et al. 1997) have allowed redshifts
to be measured for a number of GRBs (e.g. GRB 970508: Metzger et
al.~1997), providing definitive proof of their cosmological
origin. The unusually faint GRB\,980425 associated with supernova
1998bw was the first direct evidence that at least some GRB result
from the core collapse of massive stars (Galama et al. 1998).  But
other indirect evidence has come from studies of the location of GRB
in their host galaxies (e.g. Holland \& Hjorth 1999; Holland et
al. 2000; Fynbo et al. 2000; Bloom et al. 2002a; Fruchter et al. 2002;
Hjorth et al. 2002) and the types of galaxies that GRB prefer (e.g.
Hogg \& Fruchter 1999). A number of other GRB have shown deviations from a
power-law decline (e.g. GRB\,980326, Bloom et al. 1999), but these
were at high redshift and any supernova component was difficult to
study.

The very bright GRB\,011121 was detected by BeppoSAX on 2001 November
21.78288 (UT) (Piro~2001a) and its position was quickly refined to
2$'$ error radius (Piro~2001b). We began the effort to optically
monitor the error circle with the 1.3m OGLE telescope starting
$10.3\;$hours after the burst.  A possible optical afterglow (OA) was
quickly identified (Wyrzykowski, Stanek \& Garnavich 2001) as a
fairly bright ($R\approx19.1$), new object not present in the Digitized
Sky Survey image, at the position later determined by Price et al.~(2001b):
$\alpha =11^h34^m29^s\!.67,\;\; \delta = -76^\circ
01{'}41{''}\!.6\;\;{\rm (J2000.0)}$.  The source's rapid fading by
$\sim 0.5\;$mag during first $\sim3\;$hours of observations (Stanek,
Garnavich, \& Wyrzykowski 2001; Infante et al. 2001) strengthened its
likely association with the GRB. Spectra of the optical afterglow were
obtained 12~hours after the burst with the Magellan 6.5m Walter Baade
telescope and narrow emission lines from the host galaxy gave a
redshift of $z=0.36$ (Infante et al. 2001). Infante et al. went on to
note that the rapidly fading OA and the relatively low redshift of the
GRB made this burst an attractive target to search for a possible
underlying supernova. The afterglow was also detected in the infrared
(Price et al. 2001a) and radio (Subrahmanyan et al. 2001).

Late-time imaging and spectra obtained with the Magellan telescope two
weeks after the GRB suggested a slowing in the initial rapid decline
of the OA or contamination from the host galaxy of the GRB. Hubble
Space Telescope ({\em HST}) images taken two weeks after the burst and
reported in March 2002 appeared to show the burst continuing to fade
at the initial power-law rate (Bloom 2002). However, Garnavich et
al. (2002) analyzed these same images and found that the point source
was two magnitudes brighter than the extrapolation of the OA light
curve and possessed colors inconsistent with a power-law spectrum,
suggesting the presence of a supernova (SN~2001ke: Stanek et
al. 2002). Later {\em HST}\/ epochs confirmed the slow decline
consistent with a supernova (Bloom et al. 2002b; Kulkarni et
al. 2002).

Here we present our extensive data set on the optical afterglow and
the associated supernova. GRB\,011121 has been also discussed in recent
papers by Bloom et al.~(2002c), Dado, Dar \& De Rujula (2002) and
Price et al. (2002).

\section{THE PHOTOMETRIC DATA}

\subsection{$UBVRI$\/ Photometry$^9$}
\addtocounter{footnote}{1}
\footnotetext{Analysis presented here
supersedes our GCN Circulars: Wyrzykowski, Stanek, \& Garnavich
2001; Stanek et al. 2001; Infante et al. 2001; Wyrzykowski \&
Stanek 2001; Stanek \& Wyrzykowski 2001}

\begin{figure}[t]
\plotfiddle{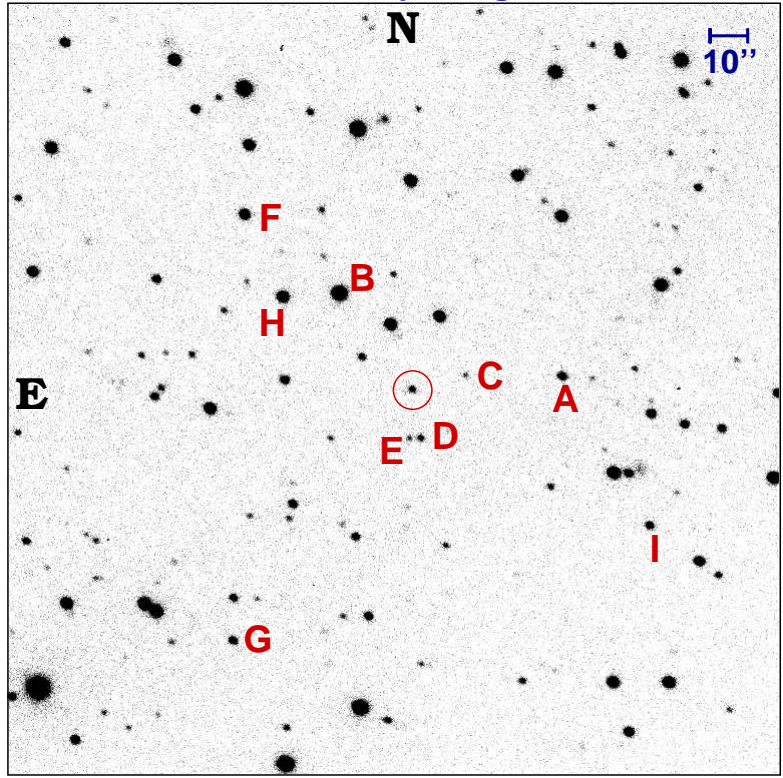}{10.cm}{0}{63}{63}{-195}{-93}
\caption{Discovery $R$-band image of the optical afterglow (circle) of
GRB\,011121, taken with the OGLE 1.3m telescope. The size of the
field is 200$''$ on a side. Also marked are photometric comparison stars (see
Table~1).}
\label{fig:finder}
\end{figure}

The majority of our photometric data were collected at two telescopes: the Optical
Gravitational Lensing Experiment (OGLE: Udalski, Kubiak \& Szyma\'nski
1997) 1.3m telescope at the Las Campanas Observatory and the Magellan
6.5m Walter Baade telescope at Las Campanas.  The OGLE telescope was
equipped with the 8k $\times$ 8k OGLE-III CCD mosaic, with a has a
scale of $0\farcs26$ per pixel.  Most of the Magellan images were taken
with the LDSS-2 imaging spectrograph in its imaging mode, with a scale
of $0\farcs378$ per pixel. Magellan images on Nov.~24, Dec.~5 and 6
were taken with the DirectCam imager with a scale of $0\farcs069$ per
pixel (unbinned). Fig.~\ref{fig:finder} shows the field containing the
OA.

Photometry obtained with the CTIO-0.9m and 4m on Nov.~22 and 23 by
Olsen et al. (2001) and Brown et al. (2001) were also added to our
data set. The CCD images were kindly provided to us and we measured the OA
magnitudes using the same methods as applied to the OGLE and Magellan data.

During the night of 2001 Nov.~22, the optical transient (OT) was
observed with the MOSAIC-II camera (chip 2) on the CTIO 4m telescope,
using $UBRI$ filters (Brown et al. 2001). That night three standard
fields (Landolt 1992) were also observed in all four colors. There
were between 4 and 6 standard stars on the chip and we have made an
independent calibration of the field stars around the GRB using this
data.  Aperture photometry was performed on the standard stars frames.
We determined the zero point for each color using average air mass
terms for CTIO.  We used DoPHOT (Schechter, Mateo \& Saha 1993) to
extract the photometry of all stars on the frames containing the OT.
We then determined an aperture correction for each frame using the
same aperture as for the standard stars. These aperture magnitudes
have been corrected for the zero point determined from the standards
stars, accounting for the difference in air mass. Color terms between
the CTIO filters and standard bandpasses were found to be small. We
estimate that our calibration is accurate to $\pm 0.02$ mag. Table~1
gives the derived magnitudes for the secondary standard shown in
Figure~\ref{fig:finder}.

\begin{figure}[t]
\plotfiddle{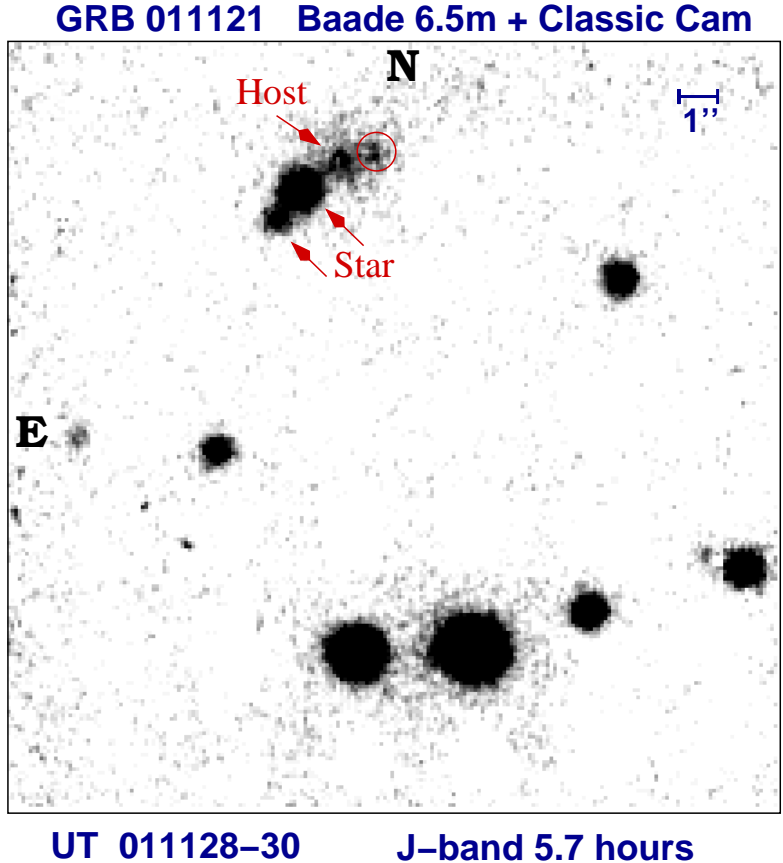}{10.cm}{0}{63}{63}{-195}{-93}
\caption{Infrared image of GRB\,011121 taken with the Magellan 'Classic Cam'.
The field is 20$''$ on a side and the small circle which marks the OA/SN has a 0.5$''$ radius.
An infrared calibration of the two bright stars in the south are given in Table~1.}
\label{fig:irchart}
\end{figure}

\subsection{{\em HST}\/ Photometry}

The Hubble Space Telescope Wide-Field/Planetary Camera~2 (WFPC2)
observed the GRB\,011121 field beginning two weeks after the burst
(GO~9180, PI:~Kulkarni). Photometry was performed on objects in the
field using a 2-pixel radius aperture. Corrections were made for
charge-transfer-efficiency (CTE) using the Whitmore-Heyer prescription
(Whitmore \& Heyer 1998) and for geometric distortion. Bright stars on the
images were used to determine the correction to bring the measurement
to an equivalent 0.5$''$ aperture. All the data were then corrected to
infinite aperture by subtracting 0.1 mag. Zeropoints on the Vega
system were taken from the WFPC2 Data Handbook and the
resulting magnitudes in the flight system bands are shown in Table~2.

To compare with ground-based $R$-band photometry the F702W magnitudes
were converted to standard $R$ using F555W$-$F702W color and
the prescription of Dolphin (2000).
Magellan $R$-band data obtained around the time of the first $HST$ epoch
are consistent with the converted F702W magnitude.

\subsection{Near-Infrared Photometry$^{10}$}
\addtocounter{footnote}{1}
\footnotetext{Analysis presented here supersedes our GCN Circular: Phillips et
al. 2001.}

 We obtained deep $J$-band images during commissioning of
the 'Classic Cam' instrument on the 6.5m Walter Baade Telescope.  The
instrument consists of a 256$\times$256 pixel NICMOS3 array with image
scale of $0\farcs094$/pixel.  Images were obtained on 2001 Nov.~29, 30
and Dec.~1 (UT) and were made up of 36, 64 and 72 individual 120s
exposures respectively. Seeing on the first night was excellent and
estimated at 0.5$''$, but this degraded to $>0.6''$ for the final two
nights. The average of the best seeing images from the three nights is
shown in Figure~\ref{fig:irchart}. The host galaxy is clearly visible
as well as two stars just to the east of the host.  We determined the
location of the OA by transferring the coordinates from the {\sl
HST\/} WFPC2 F702W images taken on 2001 Dec.~5 and noted that a faint source
was present.  Centroid determination on such a faint source results in
unreliable photometry because the aperture tends to pick the brightest
noise bump. So we fixed the position of the OA from the {\em HST}\/
data and used {\sc DaoPhot II} (Stetson 1987) and {\sc AllStar}
(Stetson 1992) to perform PSF photometry on the OA.

Three infrared standard stars (Persson et al.~1998) were observed
before and after the GRB and used to calibrate the two bright stars
south of the host galaxy (see Table~1). The ClassicCam used a $J$s
filter which is slightly narrower than the standard $J$-band, but
color corrections are expected to be small (Krisciunas et al.~2001).

In order to determine if the detections of the OA in these images were
real we ran a series of artificial star tests on the images
from each night.  We added ten artificial stars with magnitudes similar
to that of the OA to each image.  These stars were placed in regions
where the background was similar to that of the OA.  We then ran {\sc
DaoPhot II}/{\sc AllStar} on these images in exactly the same way as
we did for the original images and obtained magnitude estimates for
each artificial star.  This procedure was repeated for several sets of
artificial stars with a range of input magnitudes.  We derive a 
limiting magnitude of  $J_{lim} \approx
22.4 \pm 0.2$ for the images.  This suggests that the OA/SN was near the
magnitude limit of the data.

We subtracted the stars and used aperture photometry to measure the
brightness of the host.  The host has $J = 19.8 \pm 0.3$ in an
aperture with a radius of $4\arcsec$.  We were unable to detect any
light outside this aperture.

\section{THE SPECTROSCOPIC DATA}

Spectra of the OA were obtained with the Magellan 6.5m Walter Baade
telescope using the LDSS-2 imaging spectrograph on 2001 Nov.~22.3 UT,
$\approx 12$ hours after the burst and again on Nov.~23.3, $\approx
36$ hours post-burst.  We used a slit width of $1\arcsec$ rotated to
the parallactic angle which provided a resolution of 12~\AA\ and
coverage from 4000 to 9500~\AA . Each night of observation consisted
of two 1200s exposures on the OA as well as images of He-Ne comparison
lamps and standard stars. The spectra were bias subtracted and flat
field corrected using a normalized continuum lamp spectrum. The
extracted spectra were then wavelength and flux calibrated. The spectrum
taken on the first night showed a steep decline in flux toward the
blue, which was not consistent with the second night's spectrum.  By
calibrating stars elsewhere on the long slit we found that these too
had a blue deficit which was not accounted for by the flux
standards. We concluded that vignetting within the spectrograph had
partially blocked the shorter wavelengths and that the flux
calibration of the GRB that first night was compromised.

\begin{figure}[t]
\plotfiddle{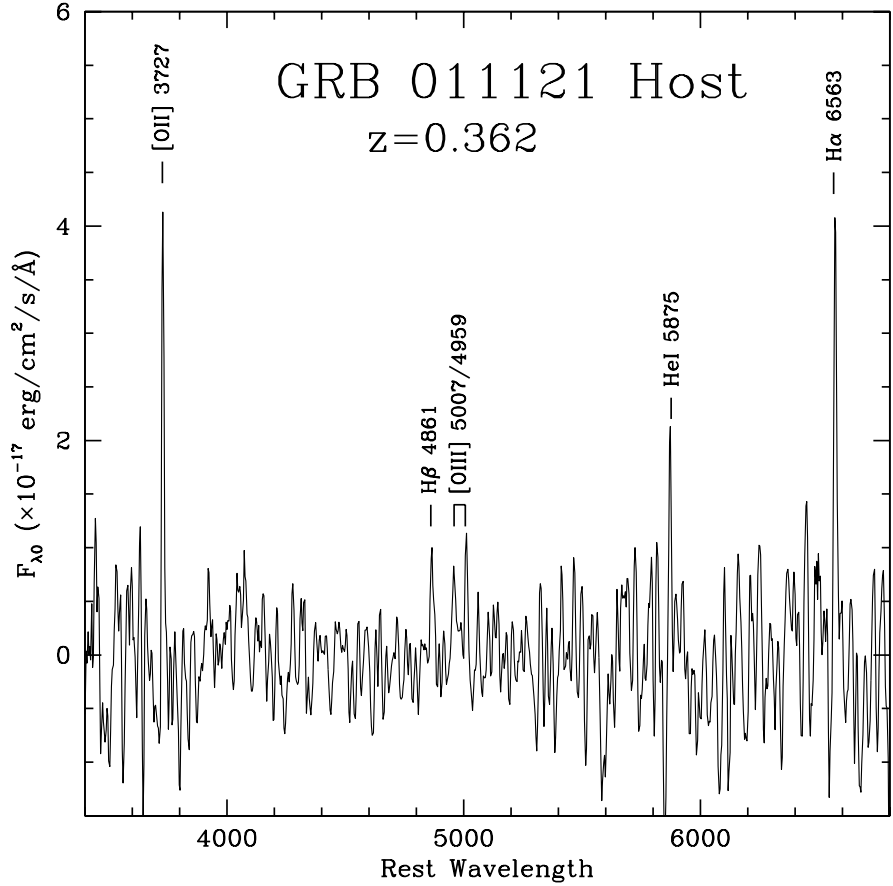}{10.5cm}{0}{63}{63}{-195}{-105}
\caption{The spectrum of the GRB plus host galaxy obtained within the
first two days of the burst. The continuum has been fit with a polynomial
and subtracted leaving only emission lines from the host.}
\label{fig:redshift}
\end{figure}

Spectra of the OA taken on Nov.~23 showed a smooth continuum with no
strong features as expected from a power-law source. Weak, narrow
emission lines were detectable in spectra from both nights. Since the
photometry showed little color change between the two nights, we used
the Nov.~23 spectrum to correct the vignetting in the Nov.~22 data and
combined all the spectra to maximize the signal-to-noise.  The
resulting spectrum with the OA continuum removed is shown in
Figure~\ref{fig:redshift}. Narrow emission features associated with
HII regions in the host galaxy are clearly present and they provide a
redshift of $z=0.362\pm 0.001$.  The line that we identify as HeI
5875~\AA\ is stronger than normally seen in typical HII regions.

Late-time spectra were again obtained with the Magellan LDSS-2
spectrograph on Dec.~7.3, $\approx 15$ days after the burst when the
supernova light dominated over the OA flux. The $1\arcsec$ slit was
rotated to include the faint star 2.5$\arcsec$ to the south east of
the OA and the host galaxy (see Figure~\ref{fig:irchart}).  The
spectra were obtained to confirm the redshift of the host galaxy as
the supernova had not been reported. Two 900s exposures were taken and
reduced in the same way as the early spectra. The seeing was good and
this permitted the two point sources to be extracted separately.
Fortunately the slit included the position of SN~2001ke and the
resulting spectrum is the first of a supernova associated with a 
cosmological GRB (Figure~\ref{fig:magellan_spec}).

\vfil\eject

\section{PROPERTIES OF THE OPTICAL AFTERGLOW}

\subsection{The GRB Temporal Behavior}

\begin{figure}[t]
\plotfiddle{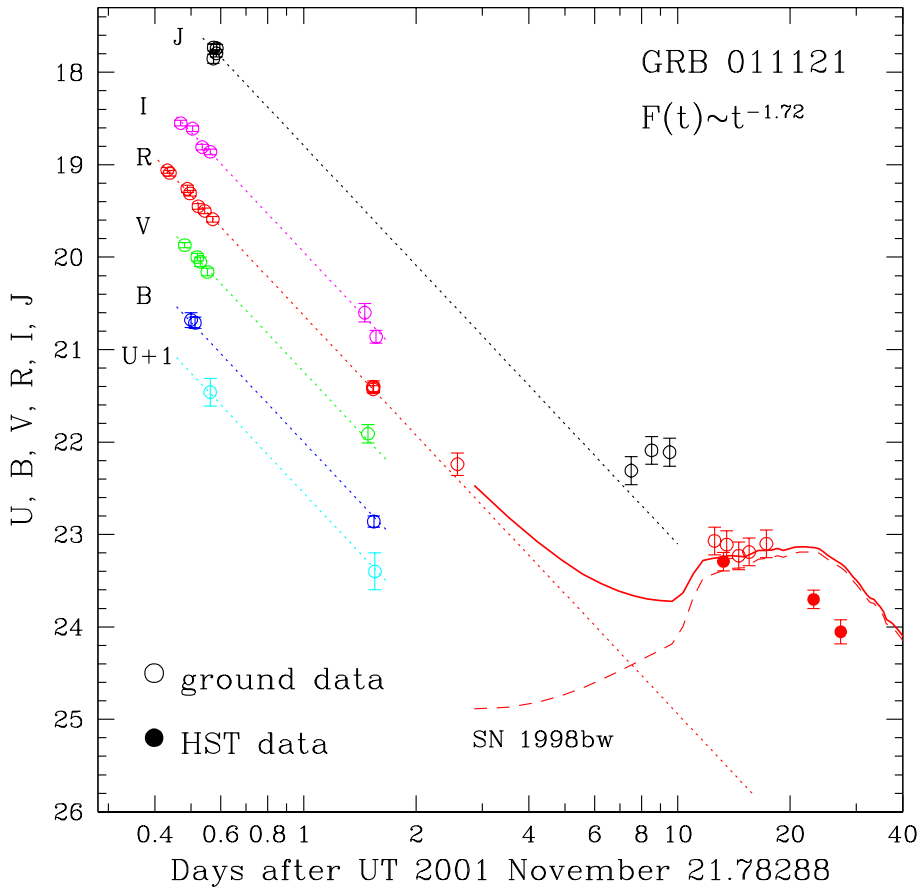}{10.3cm}{0}{63}{63}{-195}{-110}
\caption{$UBVRIJ$\/ light curves of GRB\,011121. A majority of these
data derive from this paper, with additional $UBVRI$ data from Olsen et
al.~(2001) and Brown et al.~(2001) (see Table~2) and early
$J$-band data from Price et al.~(2002). Also shown are three {\em
HST}\/ F702W epochs converted to the standard $R$-band. Dotted
lines show the OA power-law decay and the
dashed line is the light curve of SN 1998bw redshifted to $z=0.36$,
converted to the $R$-band, corrected for extinction and scaled by 0.1 mag.
The solid line shows the combination of the OA and SN~1998bw.}
\label{fig:time}
\end{figure}

We plot the GRB\,011121 $UBVRI$ light curves in Figure~\ref{fig:time}.
Majority of these data come form this paper, with additional $UBVRI$
data from Olsen et al.~(2001) and Brown et al.~(2001) which
we reduced independently to be consistent with our other photometry
(see Table~2).
In order to asses the importance of our late $J$-band data, we also
show some early $J$-band data from Price et al.~(2002). Also shown are
three {\em HST}\/ F702W epochs converted to the standard $R$-band.

The OT decays relatively quickly in early time, without a clear break
(unlike many GRBs observed since GRB\,990510: Stanek et al.~1999) or
other deviations from smooth decay, such as observed in GRB\,000301C
(e.g. Garnavich, Loeb \& Stanek 2000) or GRB\,011211 (Holland et
al.~2002).  To determine the early temporal behavior of the OA, we
have decided to fit only the data taken earlier than $2\;$days after the
burst. There are 26 such $UBVRI$ points, and as can be seen in
Figure~\ref{fig:time} a single power-law $t^{-1.72\pm 0.05}$ fit to these
early data is excellent, with the resulting $\chi^2/DOF=0.6$.

\subsection{The GRB Broad-Band Spectral Energy Distribution}

GRB\,011121 is located at Galactic coordinates of
$l=298\arcdeg\!\!.2323, b=-13\arcdeg\!\!.8633$, in a heavily reddened
direction, according to the map of Schlegel, Finkbeiner \& Davis
(1998, hereafter: SFD).  The SFD Galactic reddening towards the burst
is large, $E(B-V)=0.50$, and such large reddenings tend to be
overestimated in the SFD maps (e.g.~Stanek 1998).

\begin{figure}[t]
\plotfiddle{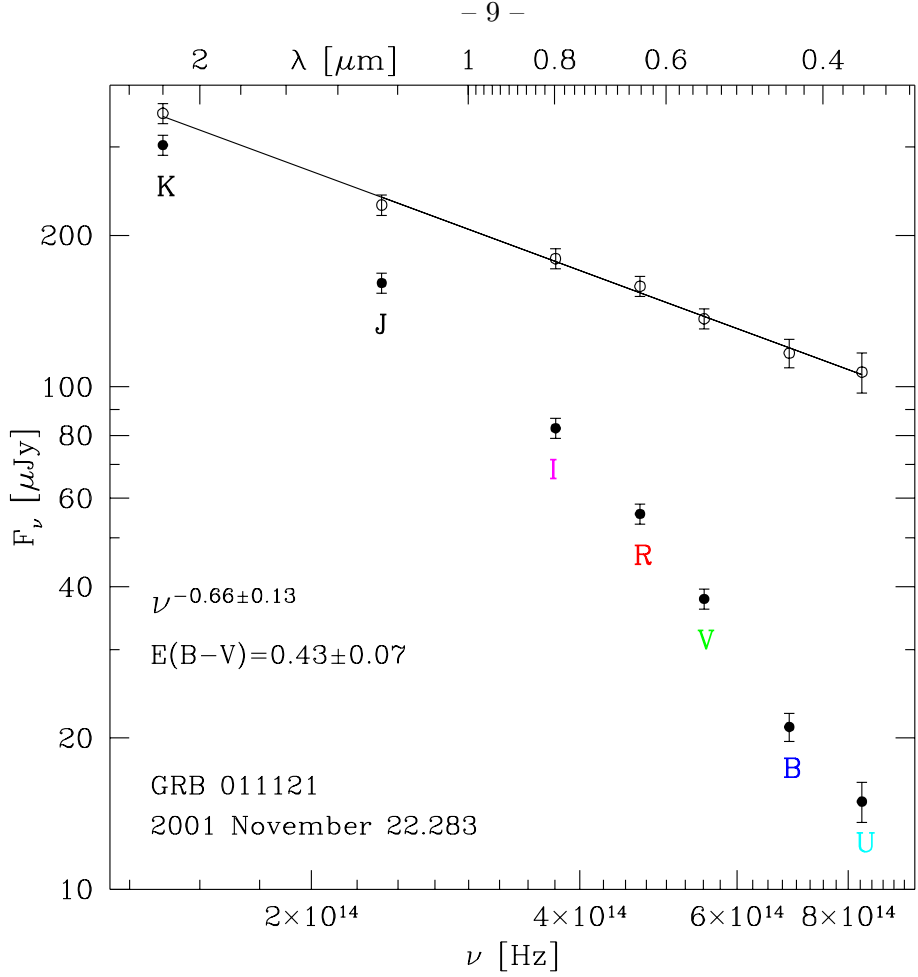}{11cm}{0}{63}{63}{-195}{-105}
\caption{Synthetic broad-band spectrum of GRB\,011121 $12\;$hours after
the burst, constructed using analytical fits shown in
Figure~\ref{fig:time}. The open symbols show the best fit power-law which
occurs for an E$(B-V)=0.43$.}
\label{fig:spectrum}
\end{figure}

To make an independent estimate of the extinction we assume that the
intrinsic energy distribution of the OA is a power-law and find the
extinction that gives the best match to a single power-law index.  We
first synthesize the $UBVRI$ spectrum from our data by interpolating
the magnitudes to a common time.  As discussed in the previous
section, the optical colors of the GRB\,011121 counterpart do not show
significant variation during the first day.  We therefore select the
epoch of November 22.283 UT ($12.0\;$hours after the burst) for the
color analysis.  We convert the $UBVRI$ magnitudes to fluxes using the
effective frequencies and normalizations of Fukugita, Shimasaku \&
Ichikawa (1995).  These conversions are accurate to about 3-4\%, so to
account for the calibration and interpolation errors we assign to each
flux a 5\% error (7\% for the $B$-band and 10\% for the $U$-band). To
get further leverage on the reddening, we add the $J$ and $K$
observations of Price et al. (2002) scaled to the fiducial time and
converted to fluxes using effective wavelengths and normalizations of
Megessier (1995).  We then determine how well a single power-law fit
the dereddened fluxes using the $\chi^2$ parameter and varying the
total extinction assuming the standard reddening curve of Cardelli,
Clayton \& Mathis (1989), as tabulated by SFD (their Table~6). We find
a best fit power-law at E$(B-V)=0.43\pm 0.07$ (one sigma) and adopt
this as the reddening to the OA. This technique lumps together Galactic
and host galaxy extinction, however, our adopted value is
one sigma lower than the SFD extinction, which implies Galactic
reddening strongly dominates over the host contribution. Our value is slightly higher
than that obtained by Price et al. (2002), most likely due to the
somewhat different calibration of the optical data.

Under the assumption that the OA spectrum is a single power-law,  the
best fit spectrum is $\nu^{-0.66\pm 0.13}$. The observed
and corrected broad-band spectra are presented in Figure~\ref{fig:spectrum}. 

\section{LATE-TIME PHOTOMETRY}

\begin{figure}[t]
\plotfiddle{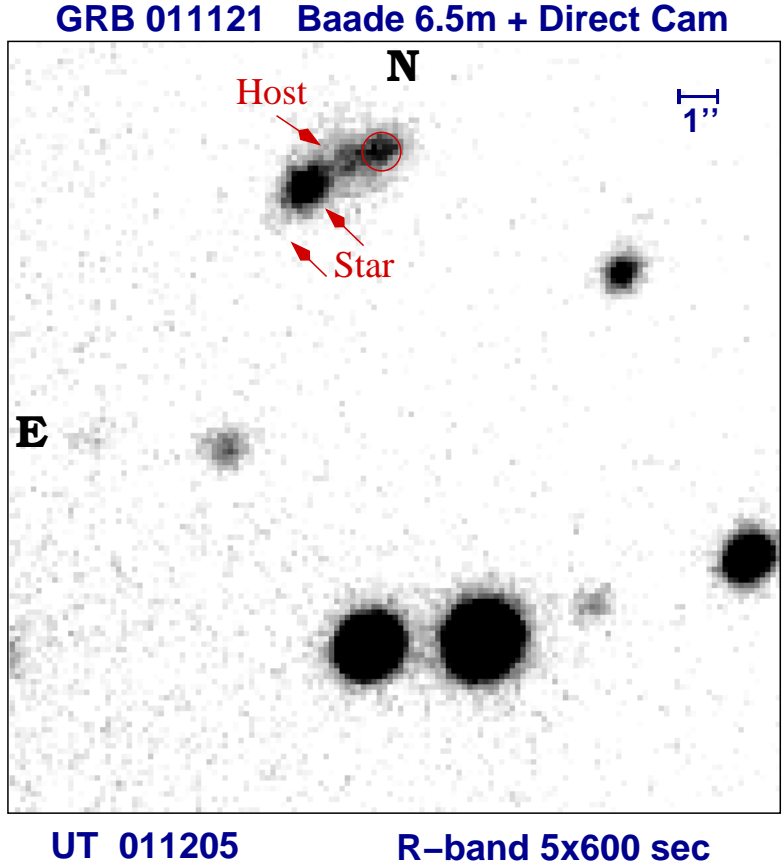}{10cm}{0}{63}{63}{-195}{-93}
\caption{$R$-band image of GRB\,011121 on 2001 Dec.~5 taken with
Magellan and Direct Cam. The circle marks the position of the OA/SN.}
\label{fig:lateimage}
\end{figure}

We searched for a late-time recrudescence in the OA by $R$-band imaging
with Magellan between 12 and 17 days after the burst.
Subtracting these images with each other revealed no significant variation
at the location of the OA suggesting that the host galaxy dominated
the emission.
When the $HST$ data was publicly released it was clear that
a point source was present which was two magnitudes brighter than the
extrapolated light curve of the OA (Garnavich et al. 2002). The position of the source was
determined from seven USNO~A.2 stars on the WFPC2 chips and found to be
11:34:29.64 $-76$:01:41.51 with an error of 0.2$\arcsec$ (Stanek et al. 2002).
This is consistent with the position of the optical afterglow determined
by Price et al. (2001b) based on the same astrometric catalog. 
 
To remove host galaxy contamination in the late-time Magellan images
we used the $HST$ data as a template. After drizzling each epoch of
F702W imaging we employed {\sc DaoPhot II} (Stetson 1987) to 
subtract the point-spread-function (PSF) of the source. 
The three epochs were then combined, smoothed
and rebinned to create a high signal-to-noise template image. Each
Magellan image was then shifted to the coordinates of the template
and the template convolved to match the PSF of the ground-based image.
The difference between the image and the template then gives the
uncontaminated brightness of the OA. The derived magnitudes are given
in Table~2 and plotted in Figure~\ref{fig:time}.

\begin{figure}[t]
\plotfiddle{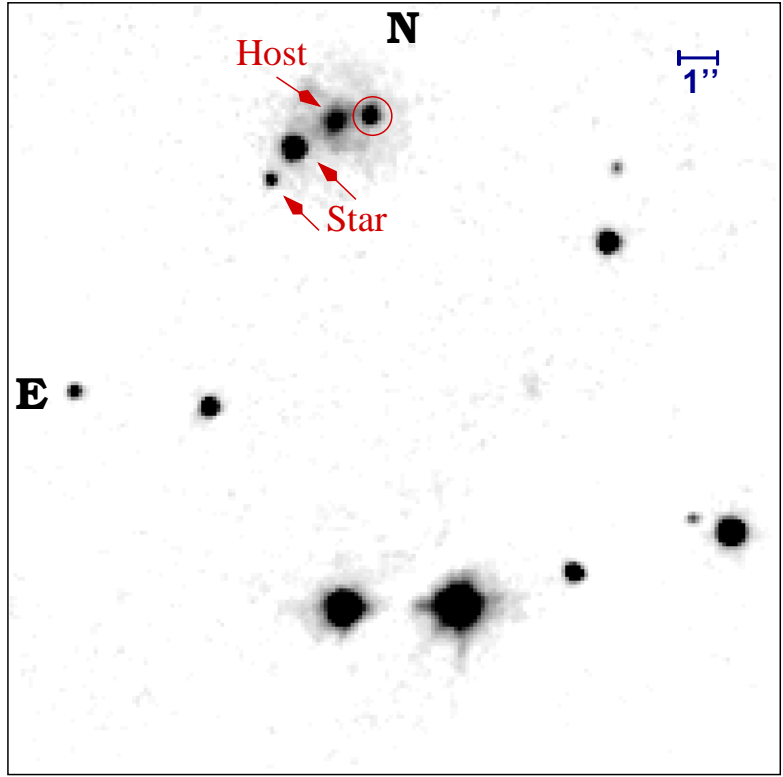}{9.5cm}{0}{63}{63}{-195}{-100}
\caption{Average of F555W, F702W and F814W {\em HST}\/ images from all
the available epochs. The extent of the host galaxy is shown as well
as a hint of spiral structure.}
\label{fig:hstimage}
\end{figure}

\section{DISCUSSION}

\subsection{GRB Properties}

For a redshift of $z=0.36$ and assuming a flat cosmology with
$\Omega_m=0.3$ and H$_0=65$ ${\rm km\; s^{-1}\; Mpc^{-1}}$, the GRB
was at a luminosity distance of 2.07 Gpc. This is the nearest OA with
a determined redshift. The GRB\,980425 associated with SN~1998bw was at
a redshift of $z=0.009$, but did not have a detected optical
afterglow.  The fluence of GRB\,011121 in the 25-100~keV band of
Ulysses was $2\times 10^{-5}$ ${\rm erg\; cm^{-2}}$ (Hurley et
al. 2001), which corresponds to an isotropic energy output of
2.7$\times 10^{52}$ erg after correction to a rest frame 20-2000 keV energy
band (Bloom et al. 2001).  This is at the low end of the isotropic
energies listed in the compilation by Frail et al. (2001) but not at
all unusual.

The $R$-band light curve of the OA shows no evidence for a break from
the time of discovery to 2.5 days after the burst. If a break occurred
later than 2.5 days then the low isotropic energy implies a jet
opening angle of $\theta > 6^\circ$. Radio observations reported by Price et al. (2002)
appear to support a late break. Correcting for the beaming
fraction, this burst would then have a total energy similar to the
`standard energy' for GRB of $5\times 10^{50}$~erg suggested by Frail
et al. (2001). 

The index of the optical light curve power-law
decay was high compared to most pre-break indices, but is typical for
post-break decay rates. The observed $\alpha =1.72$ and $\beta =0.66$
are best fit by a jet shocking a uniform density medium (ISM) and
is a poor fit to an isotropic shock into an ISM (Sari, Piran, \& Halpern 1999).
Placing GRB\,011121 on Fig.~3 of Stanek et al. (2001) suggest this OA was
observed post-break and implies the index change occurred earlier than
10~hours after the burst. An early break means a smaller opening angle
of $\theta < 3^\circ$, and an energy more than 10 times lower than
a standard burst.

The very low-redshift burst associated with SN~1998bw had an isotropic
energy well below the `standard' energy proposed by Frail et al., and
GRB\,980326, the other cosmological burst with a well-established
association with a supernova, was under-luminous compared to the
'standard' energy (Frail et al. 2001). But the numbers of events directly
associated with supernovae are too small to
link them with subluminous GRB.

If the break occurred after 2.5 days, then the steep light curve decay and flat
spectral indices observed in GRB~011121 OA are better matched to a burst
into a stratified medium instead of a uniform density gas (Chevalier \& Li 2000).
For a circumstellar
density falling as $r^{-2}$, the observed $\beta=0.66$ should 
result in a steep light curve index of $\alpha\approx 1.5$ in the slow
cooling regime.  The ISM case predicts a
more shallow $\alpha\approx 1.0$. The possibility of a circumstellar
wind has been noted by Price et~al. (2002). The presence of a
supernova at late times appears to bolster the picture of a stratified
circumburst enviroment produced by a stellar wind.

\subsection{SN 2001ke: A Supernova of a Different Color}

The late-time photometry shown in Figure~\ref{fig:time} clearly shows
a deviation from the OA power-law decay. A similar bump was detected in
GRB~980326 (Bloom et al. 1999) and attributed to a supernova although
a light echo has also been proposed to explain these
deviations (Esin \& Blandford 2000). The combination of the fading
OA and slowly rising supernova resulted in a nearly constant
brightness for the combined light during our December imaging. The
uncertainty in the brightness distribution of the host galaxy along with
the nearly constant magnitude prevented us from detecting the
supernova until the $HST$ data became available to the public.

\begin{figure}[h]
\plotfiddle{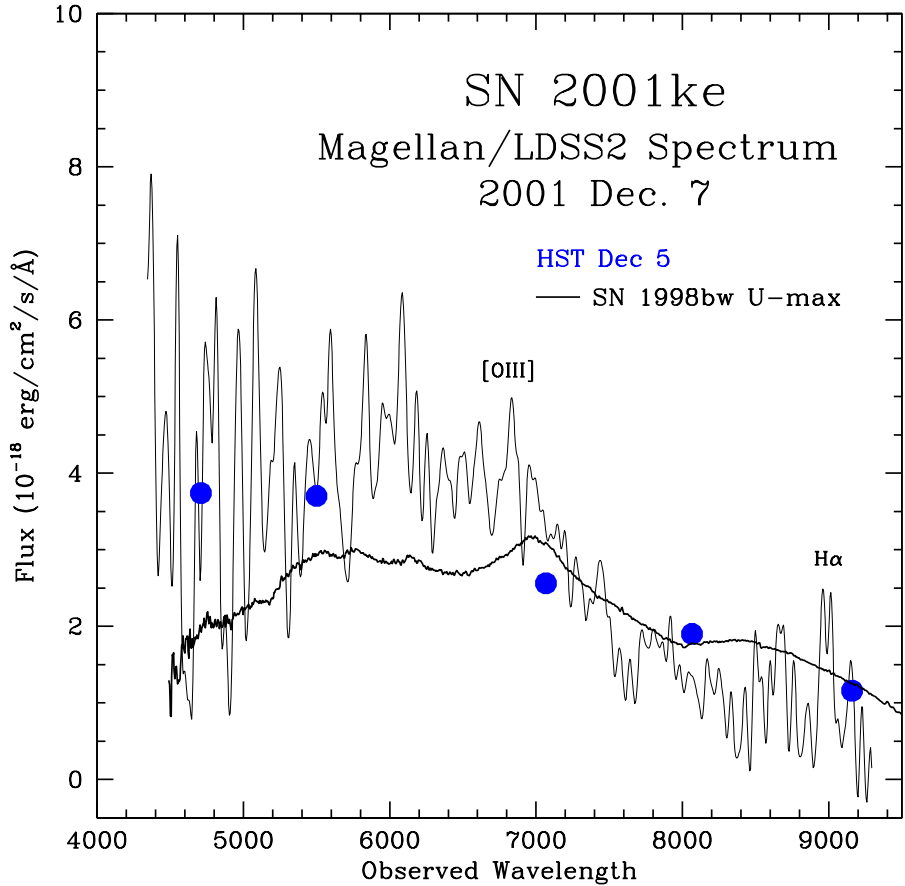}{10.3cm}{0}{63}{63}{-198}{-110}
\caption{The spectrum of supernova 2001ke taken December 7 (UT) with
the LDSS2 spectrograph on Magellan. The spectrum has been smoothed
with a 30\AA\ FWHM Gaussian kernel and corrected for extinction. For
comparison, the {\em HST}\/ broad-band flux measurements from
December~5, also corrected for extinction, are shown as large circles.
There are no obvious supernova features in the spectrum, but a comparison
with SN~1998bw at the time of $U$-band maximum light (Galama et al. 1998)
and corrected for A$_V=0.2$~mag suggests that supernova features would be
difficult to detect given the signal to noise ratio of the data}
\label{fig:magellan_spec}
\end{figure}

The first epoch {\em HST}\/ photometry of the supernova was surprising
for having a F555W-F702W color which was very blue. The
initial SN color was, in fact, similar to the OA color and this comparison
is independent of the assumed reddening. A type Ic supernova, even
an unusually energetic one like SN~1998bw is fairly red compared to the
power-law spectrum of most OA. Our spectrum,  shown in
Figure~\ref{fig:magellan_spec}, confirms the large UV flux when compared
to SN~1998bw.  The `red bump' thought to be characteristic
of a supernova is really blue in the case of GRB~011121. It is only in the later
$HST$ photometry that the supernova turns red as is characteristic of
SN past maximum.

\begin{figure}[h]
\plotfiddle{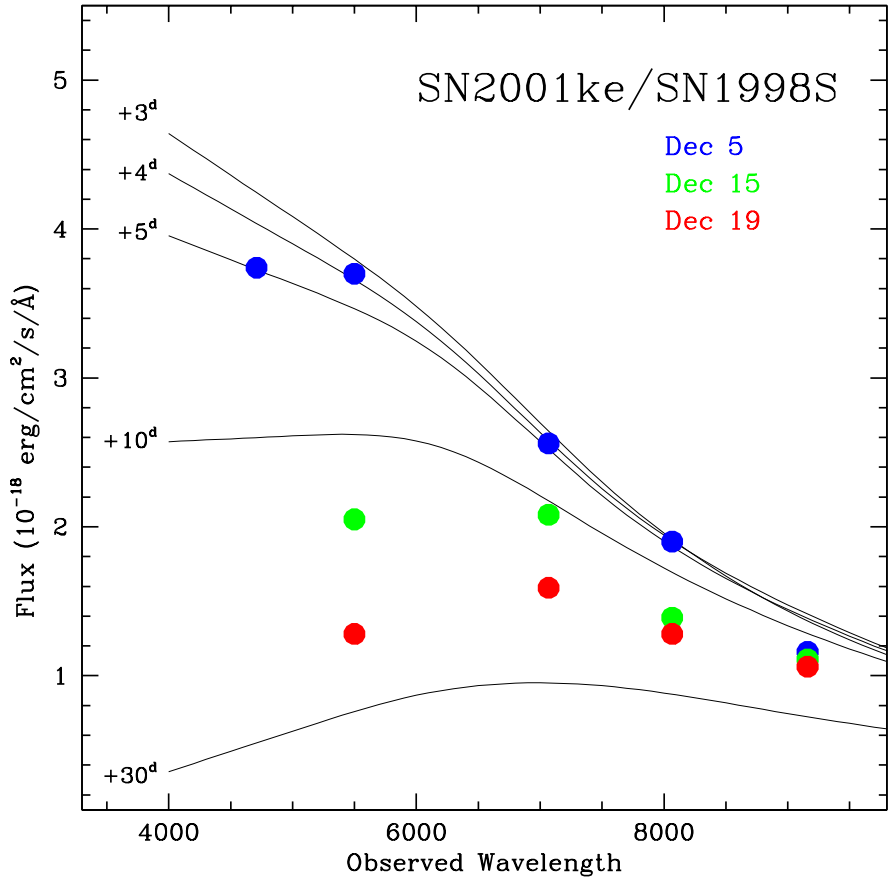}{10.3cm}{0}{63}{63}{-198}{-110}
\caption{The {\em HST}\/ photometric measurements of SN~2001ke compared with
early observations of SN~1998S. The {\em HST}\/ fluxes (circles) have been corrected for
our adopted extinction.
The fluxes for SN~1998S (lines) have been corrected for A$_V=0.12$~mag,
redshifted to $z=0.36$ and then scaled to approximately
match the early {\em HST}\/ photometry. The
times given for SN~1998S are in rest-frame days after $V$-band maximum
light.}
\label{fig:sn01ke}
\end{figure}

The colors of SN~2001ke are more typical of a type IIn events which interact with
circumstellar gas and have narrow and intermediate width hydrogen
emission lines in their spectra (Schlegel 1990). As an example we compare the spectral energy distribution
(SED) of SN~2001ke with the nearby type IIn supernova 1998S in 
Figure~\ref{fig:sn01ke}. Clearly the first $HST$ epoch is similar
to the SN~1998S SED near maximum and both supernovae redden within
a few days. Note that the infrared varies little over the observation
period as the maxima are very broad in the infrared.
SN~1998S showed Wolf-Rayet features in the
early spectra and was likely the core-collapse of a very massive star
(Lentz et al 2001).
We do not claim that SN~1998S is identical to the
GRB~011121 supernova, only that the color history of these two
events are similar and suggest circumstellar material was present.
This is consistent with the OA behavior which implied a stratified
circumstellar gas distribution.

The light curve of SN~1998S evolves much more slowly that SN~2001ke
and this is true for most type~II events with large hydrogen envelopes.
But we also note, in contrast to Bloom et al. (2002c), that the light curve
of SN~1998bw is a poor match to SN~2001ke. After correcting for
the time dilation at $z=0.362$, SN~2001ke evolves 40\%\ faster
around 5000\AA\ than 1998bw (see Figure~\ref{fig:time}) assuming the clock
started for both at the time of the gamma-ray burst. 

There is a wide range of type IIn behavior. For example, the blue color of the
type~IIb SN~1993J lasted for only a few days due to its 
unique distribution of circumstellar material (Garnavich \& Ann 1994).
Certainly SN~2001ke could have been a type Ib/c which was dominated
by circumstellar emission early on.
SN~1997cy (Germany et al. 1998) was a type IIn event which also
may have been associated with a GRB (GRB~970514).

The apparent rapid fading of SN~2001ke could also be explained if
we remove the assumption that the GRB occurred at core-collapse.
There is some evidence from X-ray spectra that metal-rich material
surround some GRB (e.g. Antonelli et al. 2000) which seems to imply that
the associated supernova exploded weeks to months before the gamma-ray emission.
In the case of GRB~011121, the time delay between core-collapse and gamma-rays 
would only be on the order of days to match SN~1998bw light curve.

Converting the SN~1998bw light curve to the $R$-band at $z=0.362$
give a peak brightness 1.3 mag less than the SN~2001ke observations.
Our adopted extinction of 1.13$\pm 0.18$ in the $R$-band suggests
SN~2001ke was similar in absolute brightness to SN~1998bw which had
a $M_V=-19.35$ (Galama et al. 1998). We estimate an absolute brightness
of SN~2001ke near maximum of $M_V=-19.2\pm 0.2$ with most of
the uncertainty due to the large extinction correction. 
 
From the $HST$ images the host appears as a face-on spiral
galaxy with the GRB/SN offset 0.88$\arcsec$ or 4.8~kpc from the center
in a low-$\Omega_m$ flat cosmology.
The [OII] flux is an good indicator of the star formation rate. After
correction for Galactic extinction and using the Balmer lines to
correct for an average host extinction, we find a [OII] luminosity
of 2.0$\times 10^{41}$~erg$\;$s$^{-1}$. We convert this to a star
formation rate using the relation given by Kennicutt (1998).
The slit did not cover the entire galaxy so the total flux is uncertain, but roughly
corresponds to a star formation rate between 3 and 6~M$_\odot\;$yr$^{-1}$.

\section{CONCLUSIONS}

We have identified the OA associated with GRB\,011121 and found its
light curve is well fitted by a single power-law decay with an index
of $\alpha =1.72\pm 0.05$ between 10 hours and $2.5\;$days after the
burst. The broad-band spectrum of the OA over this time period is well
matched by a single power-law with index $\beta =0.66\pm 0.13$ from
0.35 to 2.2~$\mu$m. We find that the reddening in the direction of the
OA is large, E$(B-V)$=0.43$\pm 0.07$. Our early spectra reveal
emission lines from the GRB host galaxy which provide a redshift
of $z=0.362\pm 0.001$. Except for the unusual GRB~980425, this is
the nearest GRB with a well determined redshift.

Our observations two weeks after the GRB reveal a source two
magnitudes brighter than expected from the OA decline. We show that
this light is likely from a supernova which was the progenitor of
GRB\,011121. We obtained the first spectrum of a supernova associated
with a cosmological GRB. The spectrum combined with colors from
{\em HST}\/ photometry show this
emission to be intrinsically blue, but becoming more red as the
supernova fades. The colors are not consistent with a typical type~Ic
and are too blue for the "hypernova" SN~1998bw. However the color can be
explained by the supernova interacting with its circumstellar
environment as was the case in SN~1998S and suggests the supernova
classification may be IIn.  Presence of circumstellar gas at the time
of the supernova implies that the GRB was not isotropic and must have
left circumstellar material for later interaction with the supernova.

The light curve of SN~2001ke appears to have peaked 10 to 12 rest-frame days
after the GRB while SN~1998bw reached maximum some two weeks
after GRB\,980425. The time evolution of SN~1998bw does not fit the
2001ke light curve well. Another type~Ic hypernova, SN~2002ap (Mazzali et al. 2002), appeared
to evolve more quickly than SN~1998bw,
however, there is no associated GRB to set the time of explosion and it appears
to be significantly less luminous than either 1998bw or 2001ke.

This early maximum combined with a blue color is surprising. Type~II supernova
rise quickly, but their hydrogen envelope traps energy and makes them
fade more slowly than observed for 2001ke. Possibly the hydrogen envelope
was small or non-existent and the circumstellar phase relatively short as was the case for
SN~1993J.  Of course, the supernova may have exploded days before the GRB
detection as has been proposed to explain metal absorption lines in
X-ray spectra of some GRB. Supernova 1998bw would fit
$R/F702W$ data well if it went off $\sim 7$ rest-frame days before the
GRB. Assuming the GRB may not signal core-collapse, we still conclude
that SN~2001ke was not identical to SN~1998bw and that a range
of supernova types can produce GRB.

GRB\,011121 is an exciting new link connecting the fields of GRBs and
supernovae, but it also represents something of a missed opportunity
to study this link in detail. Prompt and reliable reduction of the
{\em HST}\/ data for this burst (taken when the supernova component
was relatively bright) would have most likely enabled the astronomical
community to obtain significantly more information about the SN,
including better spectroscopy and perhaps polarimetry. Given that this
was the lowest redshift GRB and afterglow discovered to date (over
more than four years of intense efforts), the community should take
care that similar oversights are not repeated in the future.

\acknowledgments{We thank the BeppoSAX team, Scott Barthelmy and the
GRB Coordinates Network (GCN) for the quick turnaround in providing
precise GRB positions to the astronomical community.  We thank Knut
Olsen and Michael Brown for making their CTIO data available to us and
we thank Knut Olsen and Bohdan Paczy\'nski for useful discussions. We also thank Janusz
Kaluzny and Slavek Rucinski for taking the Nov.~24th Magellan data. 
We are grateful to F. Patat for providing spectra of SN~1998bw. 
LI wishes to thank {\em Preyecto FONDAP "Centro de Astrof\'isica"} and
{\em Proyecto Puente DIPUC} for financial support. PMG \& STH acknowledge
support from NASA grant NAG5-9364.
}

\begin{planotable}{ccccccc}
\tablewidth{10cm}
\tablecaption{Local Standard Star Magnitudes}
\tablehead{
\colhead{Star} &
\colhead{ $U$} &
\colhead{ $B$} &
\colhead{ $V$} &
\colhead{ $R$} &
\colhead{ $I$} &
\colhead{ $J$}  }
\startdata
A & \nodata  & 20.00   & 18.91\tablenotemark{a} & 18.20 & 17.58 & \nodata   \nl
B & 17.20    & \nodata & \nodata  & \nodata  & \nodata  & \nodata  \nl
C & \nodata  & 21.39   & \nodata  & 20.00    & 19.45  & \nodata   \nl
D & \nodata  & 21.03   & \nodata  & 19.26    & 18.63  & 17.83 \nl
E & \nodata  & 21.87   & \nodata  & 20.06    & 19.39  & 18.62 \nl
F &  19.61   & 19.13   & \nodata  & \nodata  & 16.70  & \nodata \nl
G &  20.32   & 20.09   & \nodata  & 18.40    & 17.77  & \nodata \nl
H &  18.39   & 18.21   & \nodata  & \nodata  & 16.22  & \nodata \nl
I &  19.98   & 19.94   & \nodata  & 18.42    & 17.83  & \nodata     
\enddata
\tablenotetext{a}{From Olsen et al. 2001}
\end{planotable}

\begin{planotable}{lcccl}
\tablewidth{15cm}
\tablecaption{GRB\,011121/SN 2001ke Photometry}
\tablehead{
\colhead{UT Date (2001)} &
\colhead{Age (days)} &
\colhead{Filter} &
\colhead{Magnitude} &
\colhead{Telescope}  }
\startdata
Nov 22.2149 & 0.4323 & $R$ & 19.06 (03) & OGLE 1.3m  \nl
Nov 22.2214 & 0.4388 & $R$ & 19.09 (03) & OGLE 1.3m  \nl
Nov 22.2519 & 0.4693 & $I$ & 18.55 (03) & OGLE 1.3m  \nl
Nov 22.2637 & 0.4811 & $V$ & 19.87 (03) & OGLE 1.3m  \nl
Nov 22.2714 & 0.4888 & $R$ & 19.26 (04) & CTIO 0.9m  \nl
Nov 22.2796 & 0.4970 & $R$ & 19.31 (03) & CTIO 0.9m  \nl
Nov 22.2824 & 0.4998 & $B$ & 20.68 (08) & OGLE 1.3m  \nl
Nov 22.2871 & 0.5045 & $I$ & 18.61 (03) & CTIO 0.9m  \nl
Nov 22.2947 & 0.5121 & $B$ & 20.71 (06) & CTIO 0.9m  \nl
Nov 22.3021 & 0.5195 & $V$ & 20.00 (04) & CTIO 0.9m  \nl
Nov 22.3057 & 0.5231 & $R$ & 19.45 (03) & OGLE 1.3m  \nl
Nov 22.3121 & 0.5295 & $V$ & 20.05 (05) & OGLE 1.3m  \nl
Nov 22.3183 & 0.5357 & $I$ & 18.81 (03) & OGLE 1.3m  \nl
Nov 22.3265 & 0.5439 & $R$ & 19.50 (03) & OGLE 1.3m  \nl
Nov 22.3357 & 0.5531 & $V$ & 20.16 (04) & OGLE 1.3m  \nl
Nov 22.3452 & 0.5626 & $U$ & 20.46 (15) & CTIO 0.9m  \nl
Nov 22.3454 & 0.5628 & $I$ & 18.86 (03) & OGLE 1.3m  \nl
Nov 22.3542 & 0.5716 & $R$ & 19.59 (03) & OGLE 1.3m  \nl

Nov 23.2400 & 1.4574 & $I$ & 20.60 (10) & OGLE 1.3m  \nl
Nov 23.2679 & 1.4853 & $V$ & 21.19 (10) & OGLE 1.3m  \nl
Nov 23.3149 & 1.5323 & $R$ & 21.41 (05) & Baade 6.5m \nl
Nov 23.3177 & 1.5351 & $R$ & 21.43 (04) & CTIO 4m    \nl
Nov 23.3201 & 1.5375 & $R$ & 21.40 (06) & Baade 6.5m \nl
Nov 23.3231 & 1.5405 & $B$ & 22.86 (06) & CTIO 4m    \nl
Nov 23.3327 & 1.5501 & $U$ & 22.40 (20) & CTIO 4m    \nl
Nov 23.3450 & 1.5624 & $I$ & 20.86 (03) & CTIO 4m    \nl

Nov 24.361 & 2.578 & $R$ & 22.24 (12) & Baade 6.5m \nl

Nov 29.31 & 7.53 & $J$ & 22.31 (15) & Baade 6.5m   \nl
Nov 30.30 & 8.52 & $J$ & 22.09 (15) & Baade 6.5m   \nl
Dec 1.30 & 9.52 & $J$ & 22.11 (15) & Baade 6.5m    \nl
Dec 4.33 &  12.55 & $R$ & 23.07 (15) & Baade 6.5m \nl
Dec 4.90 &  13.12 & $F450W$ & 24.67 (13) & {\em HST}\/    \nl
Dec 4.97 &  13.19 & $F555W$ & 23.87 (10) & {\em HST}\/    \nl
Dec 5.04 &  13.26 & $F702W$ & 23.14 (07) & {\em HST}\/    \nl
Dec 5.33 &  13.55 & $R$ & 23.11 (12) & Baade 6.5m \nl
Dec 5.83 &  14.05 & $F814W$ & 22.78 (08) & {\em HST}\/    \nl
Dec 5.94 &  14.16 & $F850LP$ & 22.78 (13) & {\em HST}\/   \nl
Dec 6.35 &  14.57 & $R$ & 23.23 (15) & Baade 6.5m \nl
Dec 7.31 &  15.53 & $R$ & 23.19 (15) & Baade 6.5m \nl
Dec 9.06 &  17.28 & $R$ & 23.10 (13) & Baade 6.5m \nl
Dec 14.82 &   23.04 & $F555W$ & 24.51 (11) & {\em HST}\/   \nl
Dec 14.88 &   23.10 & $F702W$ & 23.37 (08) & {\em HST}\/   \nl
Dec 16.64 &   24.86 & $F814W$ & 23.12 (10) & {\em HST}\/   \nl
Dec 16.75 &   24.97 & $F850LP$ & 22.73 (13) & {\em HST}\/  \nl
Dec 19.02 &   27.24 & $F555W$ & 25.02 (12) & {\em HST}\/   \nl
Dec 19.08 &   27.30 & $F702W$ & 23.66 (10) & {\em HST}\/   \nl
Dec 19.89 &   28.11 & $F814W$ & 23.21 (10) & {\em HST}\/   \nl
Dec 19.95 &   28.17 & $F850LP$ & 22.78 (13) & {\em HST}\/ 
\enddata
\end{planotable}

\end{document}